\documentclass[%
aip,
amsmath,amssymb,
reprint,%
]{revtex4-1}

\usepackage{graphicx}
\usepackage{dcolumn}
\usepackage{bm}

\usepackage[utf8]{inputenc}
\usepackage[T1]{fontenc}
\usepackage{mathptmx}
\usepackage{etoolbox}

\makeatletter
\def\@email#1#2{%
	\endgroup
	\patchcmd{\titleblock@produce}
	{\frontmatter@RRAPformat}
	{\frontmatter@RRAPformat{\produce@RRAP{*#1\href{mailto:#2}{#2}}}\frontmatter@RRAPformat}
	{}{}
}%
\makeatother
\begin{document}
	
	\preprint{AIP/123-QED}
	
\title[MDS Room Acoustics]{Non-verbal  Perception of Room Acoustics using Multi Dimensional Scaling Method}
\author{Leonie Böhlke-Grabe}
\affiliation{Taubert und Ruhe GmbH, 25421 Pinneberg, Germany}
\author{Tim Ziemer}
\affiliation{Institute of Systematic Musicology, University of Hamburg, Germany}
\affiliation{KREBS + KIEFER Ingenieure, 20457 Hamburg, Germany}
\author{Rolf Bader}
\affiliation{Institute of Systematic Musicology, University of Hamburg, Germany}
\email{R\_Bader@t-online.de}

\date{\today} 

\begin{abstract}
Subjective room acoustics impressions play an important role for the performance and reception of music in concert venues and auralizations. Therefore, room acoustics since the 20th century dealt with the relationship between objective, acoustic parameters and subjective impressions of room acoustics. One common approach is to correlate acoustic measures with experts' subjective ratings of rooms as recalled from their long-term memory, and explain them using acoustical measures. Another approach is to let listeners rate auralized room acoustics on bipolar scales and find objective correlates. In this study, we present an alternative approach to characterizing the subjective impressions of room acoustics. We concolve music with binaural room impulse response measurements and utilize Multi Dimensional Scaling (MDS) to identify the perceptual dimensions of room acoustics. Results show that the perception of room acoustics has $5$ dimensions that can be explained by the (psycho-)acoustical measures echo density, fractal correlation dimension, roughness, loudness, and early decay time.\end{abstract}

\maketitle


\section{\label{sec:1} Introduction}
Room acoustics, particularly important for music venues \cite{beranek}, must be related to music perception in music spaces. Such subjective room acoustics refer to perceptual listening impressions \cite[chap. 6]{ziemerbook}. As with timbre, these impressions are multidimensional, where most studies find subjects to discriminate timbre in three perceptual dimensions (see, e.g., \cite{grey}; for a review, see \cite[Chap. 11]{badernonlinearities}). Books on architectural acoustics and spatial audio use various terms to characterize subjective room acoustics, such as \emph{reverberance}, \emph{spaciousness}, \emph{liveness}, \emph{spatial impression}, \emph{listener envelopment}, \emph{intimacy}, or \emph{apparent source width}\cite{beranek}\cite[chap. 6]{ziemerbook}. 

Several approaches have been developed to capture subjective room acoustics systematically. The Spatial Audio Quality Inventory (SAQI) \cite{saqi} provides a comprehensive vocabulary for room acoustics and virtual acoustics that includes three to ten terms assigned to the categories \emph{timbre}, \emph{tonalness}, \emph{geometry}, \emph{room}, \emph{time behavior}, \emph{dynamics}, \emph{artifacts}, and \emph{general}. Its room acoustical counterpart, the Room Acoustical Quality Inventory (RAQI), suggests items or factors, such as \emph{quality}, \emph{strength}, \emph{reverberance}, \emph{brilliance}, \emph{irregular decay}, \emph{coloration}, \emph{clarity}, \emph{liveliness}, or \emph{intimacy}, plus four additional terms. Beranek suggested $16$ attributes and respective measurement parameters to estimate the 'quality' of concert halls \cite{Beranek1962}.

Criticizing the amount of rooms used in listening test, a minimum of $60$ rooms was proposed, along with new methods of virtual room acoustics and 3D recordings \cite{Weinzierl2015}. Additionally, the influence of subjective preferences or the music played in the respective rooms was critically discussed, and more objective means are demanded. In a Multidimensional Scaling (MDS) experiment \cite{Hawkes1971}, six dimensions were found and associated with \emph{balance and blend}, \emph{resonance}, \emph{intimacy}, \emph{brilliance}, \emph{proximity}, and \emph{definition}. Additionally, a factor analysis was conducted, yielding similar results \cite{factoranalysis}. The stimuli were taken from different artificial reverberations used in the Royal Festival Hall in London. The subject ratings were done on bipolar scales like 'brilliant -- dull' or 'distant -- close'. Also, a pairwise comparison was used to identify latent classes\cite{lokki}. Here, it was found that one group prefers sounds that are proximate, enveloping, and warm, while another prefers clarity. All these subjective attributes can be associated with one or more room acoustical attributes, like Early Decay Time (EDT), RT$_{20}$, strength G, lateral sound level $L_J$, and early lateral energy $J_{LF}$.


\begin{figure}[ht]
\includegraphics[width=\linewidth]{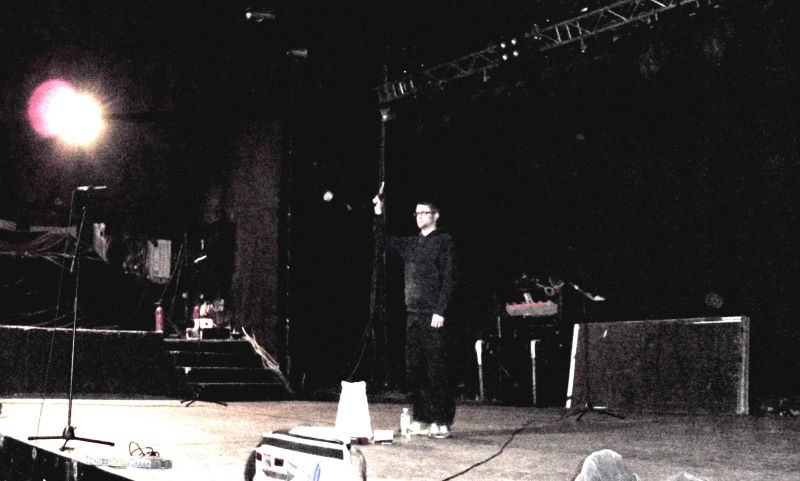}
\caption{\label{pic:niko}{Room Impulse Response (IR) measurement setup, using an alarm pistol, shot on a music venue stage.}}
\end{figure}



 
Most of these previous studies used mainly verbalizations to characterize room acoustic, most often adjectives. Such verbalizations were additionally most often presented as bipolar scales. Thereby, the relation between the physics of the rooms and the subjective ratings makes a detour via language, which introduces an additional bias. The present study avoids such verbalizations by only asking for similarities between musical stimuli, performing a Multidimensional Scaling analysis, and correlating the perceptual dimensions with psychoacoustic and room acoustic parameters. Although parameters with high correlation can then be verbalized in a last step, like associating a spectral centroid with perceived brightness or a calculated roughness with perceived roughness, this last step of verbalization is not part of the listening test or the analysis part and might be omitted entirely. 
	
The  MDS method used in the present study was derived from listening tests in timbre analysis (for a review see \cite[Chap. 11]{badernonlinearities}) and has several advantages. Firstly, it omits the blurredness of language with the subjective interpretation of adjectives, as discussed above. Secondly, it omits the bias of a finite list of adjectives presented by the experiment beforehand, where some important adjectives might not be present. Thirdly, it omits bipolar scaling, which enforces a subject's judgment in a bipolar way. Bipolar scales are a bias, as subjects might find both poles to be strongly or weakly present in the sound at the same time. So, e.g., a subject might find a musical sound to be bright and dark at the same time when, e.g., a strong bass instrument is accompanying a bright lead instrument. The cause of such judgments is found in auditory scene analysis, where listeners extract several sonic objects in a sound, making judgments on each of them. In musical terms, this can be a polyphonic texture where each melodic line is perceived separately; it might also be a lead melody accompanied by instruments playing chords. Yet, a fourth advantage is omitting terms like quality from a rating. Although ratings might sometimes be important for listeners, such ratings strongly depend on the musical pieces played and the venue built for each piece. Rooms with variable acoustics point to such complex uses.

Previous studies often used clusters to extract main features for room acoustic perception. This includes the search for orthonormal parameters. Indeed, in psychology and psychoacoustics all perceived dimensions are confounded and never orthonormal. The best-known example might be the perceived pitch of a sinusoidal frequency rising when increasing its amplitude while keeping its frequency constant. Therefore, none of the psychoacoustic parameters are expected to be perfectly orthonormal. In cases where more than one parameter is correlating significantly with a perceptual dimension, there is no explicit need to decide on one of them, and additional reasoning needs to be taken into consideration, as discussed below in the results section in detail.
	
One disadvantage of the proposed MDS method is the restriction of the number of sounds possibly used. As proximity estimations need each sound to be compared to all other sounds, already nine sounds used in this paper need $36$ pair judgments. Increasing the amount of sounds is very soon overburdening subjects' concentration. Also, the choice of the dry signal the room impulse responses are convolved with is expected to lead to different results. Therefore, this study is a first step towards non-verbal-based investigations of perceived room acoustics, which already leads to unexpected results with additional psychoacoustic parameters fitting perception better than most conventional parameters.

\section{Method}
\label{method}
Room impulses of three Hamburg concert halls were used. These were convolved with a dry musical piece, resulting in the stimuli presented to subjects in a Multidimensional Scaling (MDS) listening test. The room impulse responses, as well as the convolved music, were analyzed using room acoustic and psychoacoustic parameters. Finally, the dimensions resulting from the MDS analysis were correlated with these acoustic and psychoacoustic parameters to explain the dimensions found when subjects discriminate between these different rooms or spatial perception in the musical piece.

\subsection{Room Impulse Responses}
\label{rir}
Room impulse responses were measured in three concert venues using a blank pistol on the stages of the unoccupied rooms, as illustrated in Fig. \ref{pic:niko}. These were recorded at typical listening positions using a HEAD acoustics HSU III.2 dummy head and several Behringer ECM-8000 measuring microphones. 

The three venues were Markthalle, Fabrik, and Laiszhalle Hamburg. The first two venues are mostly used for popular music concerts and nightclub events, the latter is a location for symphony music and popular music concerts with a seated audience. In each venue, we recorded three different source-receiver constellations and selected typical ones for the listening test.

\subsection{Preparation of Stimuli}
As a music stimulus, we recorded a banjo piece in an anechoic chamber in the Institute of Systematic Musicology. This near field mono recording was convolved with the binaural room impulse responses as recorded with the dummy head. The resulting stereo sounds were played back via loudspeakers in the lecture room at the Institute to $11$ participants.

Listening experiments with music that has been convolved with room impulse responses are quite common \cite{lokki}, as listeners tend to have more experience with music than with impulse responses.

\subsubsection{Multidimensional Scaling (MDS) Listening Experiment}
The participants evaluated the perceptual proximity on a Likert scale from $1$ (completely similar) to $7$ (completely dissimilar). With $9$ items, this means $8+7+6+5+4+3+2+1=36$ comparisons. To avoid any bias, we did not present examples of music with very similar or very dissimilar room acoustics. Consequently, we only consider the dissimilarity ratings as data in an ordinal scale, not on an interval level.

We embed the dissimilarity ratings through nonmetric Multi Dimensional Scaling (MDS). As MDS is a dimensional reduction, higher embedding dimensions lead to a better embedding of perception ratings into the embedding. Higher embeddings normally converge to a remaining, unexplainable data noise. For the present MDS test, we considered up to six dimensions reasonable.

The MDS is carried out for each individual, and then aggregated over all $11$ participants, as suggested in \cite[p. 973]{brosius}. We consulted STRESS 1 and R-squared (RSQ) criterion to embed the similarity ratings in a space with $1$ to $6$ dimensions. The orientation values suggested by Kruskal \cite[p. 3]{kruskal1964} are listed in Table \ref{tab:Model_Quality}. They characterize the relationship between STRESS 1 and model quality (fit). STRESS 1 is calculated individually for each participant. The Root Mean Square (RMS) of STRESS 1 is chosen for the aggregated model, as suggested in \cite[p. 175]{Schiffman}.

\begin{table}
    \centering
    \begin{tabular}{cc}
        STRESS 1 & Fit\\
        \hline\\
       $0.2$    & Poor\\
       $0.1$    & Fair\\
       $0.05$   & Good\\
       $0.025$  & Excellent\\
       $0$      & 'Perfect'\\
    \end{tabular}
    \caption{Rough interpretation of Stress 1 as model fit quality.}
    \label{tab:Model_Quality}
\end{table}

RSQ values indicate the squared correlation between distances and disparities, whereby the values also lie between $0$ and $1$. Here, the higher the value, the better the fit. We calculate RSQ for each individual, and take the mean value to evaluate the model that takes all $11$ participants into account.

\subsubsection{Test conditions and subjects}

\subsection{Acoustic and Psychoacoustic Parameters}
\label{parameters}
As possible candidates to explain the dimensions of room acoustical perception, we chose established parameters from the field of psychoacoustics, room acoustics, and some unconventional parameters that have been found useful in previous timbre studies. Room acoustic parameters used are reverberation time (RT), Early Decay Time (EDT), and loudness (N), parameters found important for timbre discrimination included are spectral centroid (C), sharpness (S), roughness (R), fluctuation strength (F), and tonality (K), as well as the fractal correlation dimension (D)\cite{badernonlinearities,badermusic,BaderFracLigeti,BaderFracGuitar,BaderFracInstruments,borgo}, and a measure for the event density, a rather uncommon but highly effective parameter, the Echo density (E) suitable for room acoustics \cite{BaderGenre}. As two RT and two roughness estimations are used, a total of twelve psychoacoustic parameters were calculated. We utilized Apollon \cite{apollon} software suite for timbre parameter calculations. Note that the EDT and RT are calculated from the room impulse response directly, whereas the other parameters are calculated from the music piece after it has been convolved with the respective impulse response. As these parameters have different formulations, we briefly discuss their definition in the present study in the following sections.

\subsubsection{Spectral Centroid ($C$)}
The spectral centroid $C$ is the center of gravity of a spectrum, where the sum of amplitudes above and below this center is equal, and calculated as
\begin{align}
C = \frac{\sum_{i=0}^N f_i A_i}{\sum_{i=0}^N A_i} \ .
\end{align}
This corresponds to psychoacoustic brightness perception \cite{beauchampcentroid,saqi}.

\subsubsection{Roughness ($R$)}
Roughness calculations have been suggested in several ways (for a review, see \cite{schneiderrough,badernonlinearities}). We utilized two algorithms, calculating the beating of two sinusoidals close to each other (Helmholtz\cite{helmholtz}, Helmholtz/Bader\cite{schneiderrough}, Sethares\cite{sethares1993}), or integrating energy in critical bands on the cochlear (Fastl\cite{zwicker}, Sottek\cite{sottek1993}). The former has been found to work very well with musical sounds, the latter with industrial noise. In this paper, both, a modified Helmholtz/Bader algorithm and the algorithm of Sottek are used.

With the Helmholtz/Bader algorithm, like the original suggestion of Helmholtz, a maximum roughness appears at a frequency difference of $33$ Hz between two sinusoidals. As Helmholtz did not provide a mathematical formula, according to his verbal descriptions, a curve of the amount of roughness $R_n$ is assumed between two frequencies with distance $df_n$ which have amplitudes $A_1$ and $A_2$ like
\begin{equation}
R_n = A_1 A_2 \frac{|df_n|}{f_r e^{-1}} e^{- |df_n|/f_r} \ .
\end{equation}
with a maximum roughness at $f_r = 33$ Hz. The roughness $R$ is calculated as the sum of all possible sinusoidal combinations like
\begin{equation}
R = \sum_{i=1}^N R_i \ .
\end{equation}

The only difference between the algorithm used in Apollon and that described in \cite{schneiderrough} is the precision with which the frequencies are calculated. To arrive at very precise values in \cite{schneidertimbre}, a wavelet analysis is performed, allowing for an arbitrary precision of frequency estimation. As this is very computationally expensive, in the present study, the precision of Fourier analysis described above is used. In \cite{schneiderrough}, the research aim was to determine the perceptual differences between tuning systems such as Pure Tone, Werkmeister, Kirnberger, etc. in a Baroque piece of J. S. Bach. The present analysis is not aimed at such subtle differences, but rather at estimating the overall roughness, similar to other room acoustical studies \cite{saqi}.

\subsubsection{Sharpness ($S$)}
Perceptual sharpness is related to the work of Bismarck \cite{bismarck} and followers \cite{aures1985a,aures1985b,fastl}. It corresponds to narrow frequency-band energy. According to \cite{fastl}, it is measured in acum, where $1$ acum is the narrow-band noise within one critical band around $1$ kHz at $60$ dB. Sharpness increases nonlinearly with frequency. If a narrow-band noise increases its center frequency from about $200$ Hz to $3$ kHz, sharpness increases slightly, but above $3$ kHz strongly, according to perception that very high narrow-band sounds have strong sharpness. Still, sharpness is mostly independent of overall loudness, spectral centroid, or roughness, and therefore qualifies as a parameter on its own \cite{saqi}. 

To calculate sharpness, the spectrum $A$ is integrated with respect to $24$ critical or Bark bands, as we are considering narrow-band noise. With loudness $L_B$ at each Bark band $B$ sharpness is
\begin{equation}
S = 0.11 \frac{\sum_{B=0}^{24 Bark} L_B g_B B}{\sum_{B=0}^{24 Bark} L_B} \ \text{acum} ,  
\end{equation} 
where a weighting function $g_B$ is used strengthening sharpness above $3$ kHz \cite{Peeters2004} like 

\begin{equation}
g_B = \left\{\begin{array}{ll} 1 \text{ if } B < 15 \\ 0.066 e^{0.171 B} \text{ if } z \geq 15 \end{array} \right.
\end{equation}

\subsubsection{Loudness ($L$)}
Although several algorithms of sound loudness have been proposed \cite{fastl,sottek}, for music, still no satisfying results have been obtained \cite{ruschkowski2013}. Most loudness algorithms aim for industrial noise, and it appears that musical content considerably contributes to perceived loudness. Also, loudness perception differs  significantly between male and female subjects due to the different constructions of the outer ears between the sexes. Therefore, a very simple estimation of loudness is used, and further investigations in the subject are needed. The algorithm used is
\begin{equation}
L = 20 \log_{10} \frac{1}{N}\sqrt{\sum_{i=0}^N \frac{A_i^2}{A_{ref}^2}} \ .
\end{equation}
This corresponds to the definition of decibel, using a rough logarithm-of-ten compression according to perception and multiplying with $20$ to arrive at $120$ dB for a sound pressure level of about $1$ Pa. Of course, the sounds played to listeners differed in SPL to those present at the recording location in the music venue. Still, the SPL relations between the IR are preserved.

\subsubsection{Fluctuation Strength ($F$)}
\label{fluctuation strength}
The psychoacoustic parameter fluctuation strength was calculated using the Sottek model \cite{sottek1993,sottek1994}. It detects the modulation depth of a sound for low modulation frequencies around $4$ Hz, as higher modulation frequencies lead to roughness. It is calculated using Bark bands and is measured in the unit of \emph{vacil}, where $1$ vacil is for a $1$ kHz sinusoidal at $60$ dB, amplitude modulated at a depth of $100$ \% with a modulation frequency of $4$ Hz. 

\subsubsection{Tonality ($T$)}
\label{tonality}
Tonality or tonalness is another distinct auditory sensation \cite{saqi}. A noise is considered tonal, if individual tones or tonal components clearly stand out. In this sensation, again, the processing of sounds in the various critical frequency bands plays an important role, where the critical bandwidths of the frequency bands are crucial and determine the audibility of tonal components \cite[p. 70]{genuit}.

For the calculation of the tonality, different models have been proposed \cite{terhardt,aures1985a,aures1985b}, according to which the tonal part of a sound is separated from the noise content. The level of the tonal part is then calculated after making corrections for masking effects, bandwidth, frequency, and loudness (according to \cite{maschke} p. 18).

If it is assumed that two cases can be distinguished with regard to the tonal components, namely, on the one hand, dominant frequencies or spectral lines and, on the other hand, dominant narrow-band noise components, then the tonality can be calculated as follows: In a first step, the Fourier spectrum of a signal is calculated and scanned for dominant spectral lines by using the formula
\begin{equation}
    L_{i-1} < L_{u} \geq L_{i+1} \ .
\end{equation}
Here, $L_i$ denotes the level of the $i$th dominant frequency in the spectrum, $L_{i-1}$ that of the next lower, and $L_{i+1}$ that of the next higher frequency. In a second step, it is tested whether a found, dominant frequency represents a tonal component by using the following formula

\begin{equation}
    L_i - L_{i+j} \geq 7 \text{ dB}, j = -1,-2,+1,+3 \ .
\end{equation}

If each of these conditions is fulfilled, it is assumed that the considered group of seven spectral lines with the indexes $i-3$, $i-2$, $\ldots i+3$ represents a tonal component\cite[p. 18]{maschke} \cite[p.681]{terhardt}.

In a third step, remaining narrow-band noise components of relatively high intensity are detected whose bandwidth is smaller than a critical bandwidth at this location. They are considered a tonal component when the sound pressure level in the adjacent frequency groups is at least $7$ dB lower. Such components are also removed from the spectrum, leading to a separation of the tonal and noise components.

The tonality of the individual spectral components depends on their frequencies, level, and bandwidth, whereby each of them is subject to a weight, $w_1(\Delta z_i)$ for bandwidth, $w_2(\Delta L_i)$ for level, $w_3(\Delta f_i)$ for frequency, and $w_{GR}$ as the level relation of noise to tonal components, resulting in a calculation for tonality like
\begin{equation}
    K = c \  w_{GR}^{0.79} \left(\sqrt{\sum_{i=1}^N \left(w_1(\Delta z_i)  w_2(\Delta L_i) w_3(\Delta f_i)\right)^2}\right)^{0.29} \ .
\end{equation}
Here $\delta L_i > 0$ and $c$ is a constant that normalizes the result to the reference sound. The unit of tonality is \emph{tu}, which stands for tonality unit. The reference sound used to determine the tonality of $1$ tu is a $1$ kHz sine tone at a sound pressure level of $60$ dB.

\subsubsection{Reverberation Time (RT)}
\label{reverberation}
In this study, the $RT_{20}$ and $RT_{30}$ reverberation times are calculated from the measured room impulse responses by extrapolating the time interval between a $-5$ dB and a $-25$ or $-35$ dB level drop, respectively, using Schroeder's backward integration method \cite{schroeder}.

\subsubsection{Early Decay Time (EDT)}
The EDT is a similar measure, only for a level drop from $-0.1$ dB to $-10.1$ dB, measuring the very first decay of early reflections, which can be considerably different from that of the following tail of the impulse response\cite{ziemerbook}.

\subsubsection{Fractal Correlation Dimension ($D$)}
\label{fractal}
The fractal correlation dimension has often been used to calculate the event density of musical pieces or sounds\cite{badermusic,BaderFracLigeti,BaderFracGuitar,BaderFracInstruments,borgo}. It uses a pseudo-phase space to obtain a multi-dimensional embedding of a single, discrete time series $x_i$ with $i=1,2,3..., N$ and a delay variable $\delta$, where for each time point $i$, a new vector $y_i$ is built like
\begin{equation}
	y_i = \{x_i, x_i + \delta, x_i + 2 \delta, x_i + 3 \delta,...,x_i
	+ M \delta \} \ ,
\end{equation}
with the integer $\delta > 2$ and integration depth $M$. All $N-M$ vectors built a cloud of points in the $M$-dimensional space. All distances $r_{i,j}$ between point $y_i$ and $y_j$ are forming a distance histogram $C(r)$ when using the Heaviside function $H$ for sorting all distances within a hypersphere of radius $r$ like
\begin{equation}
	C(r) = \frac{1}{N^2} \sum_{i \neq j} H(r-|y_i - y_j|) \ .
\end{equation}
The Heaviside function $H$ is defined as
\begin{equation}
	H(x) = \left\{%
	\begin{array}{ll}
		0, & \hbox{for x} < 0 \\
		1, & \hbox{for x} \geq 0 \\
	\end{array}%
	\right .
\end{equation}
Now, the correlation dimension is the region of the plot $\log C(r)$ vs. $\log r$ with a constant slope like
\begin{equation}
	D_C = \lim_{r \rightarrow 0} \frac{\ln C(r)}{\ln r} \ .
\end{equation}
$D_C$ is counting the amount of dynamical subsystems in the time series. This is straightforward in terms of musical applications, as one dynamical subsystem is one harmonic overtone spectrum, no matter how many partials are present in this spectrum. Each additional harmonic overtone series, which is inharmonic to all other ones present in the sound, raises $D_C$ by one. Also, large amplitude fluctuations can lead to a rise in $D_C$.

So we can conclude that within one operation, the amount of pitches and large fluctuations are counted and stored in the value of $D_C$. These parameters are the first to come to mind in terms of event density. If an inharmonic or noisy sound is present, $D_C$ rises with each new strong inharmonic partial. Still, an amplitude threshold is present, where very weak inharmonic partials are not accounted for. In most cases, this is an advantage as the noise floor would be included in the calculation, which we are not interested in. Still, if the noise is getting louder, $D_C$ rises tremendously.

\subsubsection{Echo Density ($E$)}
\label{echo}
An echo density was proposed to estimate the amount of echos present in a sound \cite{BaderGenre}. If a sinusoidal wave of frequency $f$ is present at a position in a room, its phase repeatedly and continuously cycles from $0$ to $2 \pi$. If this wave is reflected at a wall in that room and returns to that position at a certain point in time, the phase of the returning wave is very likely different from the phase of the wave already present at that position at the time of return. Therefore, the phase of the sinusoidal wave at the respective position will make a sudden jump or shift to a phase of both waves superposing one another. The echo density  $E(f,t)$ then depends on frequency and time like
\begin{equation}
E(f,T) = |A(f,T)| \frac{\partial^2 \phi(A(f,T))}{\partial t^2} / (2 \pi) \ ,
\end{equation}
where $|A(f,t)|$ is the amplitude of the respective wave at frequency $f$ and adjacent time windows $T$, and $A(T)$ is calculated using a Hanning-windowed Fourier Transform. As the analyzing frequencies in the Fourier Transform fit into the time window size $\mathrm{d}T$ as integer multiples of $2\pi$ of the frequencies respective phases, for a continuous analyzed sinusoidal adjacent windows have a constant phase and therefore
\begin{equation}
    \frac{A(f,T_2) - A(f,T_1)}{dT} = \frac{\partial A(f,T)}{\partial t} = 0 \ .
\end{equation}
Still, a constant rise or fall of phases has a non-zero first derivative. To ensure that only sudden phase shifts are detected and not the continuous travel of the phase through time, a second derivative of the phase over time is used.

Summing over frequencies and time points, an echodensity of the piece of music, a sound, or an impulse response is
\begin{equation}
    \textbf{E} = \sum_{f,T} E(f,T) \ .
\end{equation}
Echos are not only present in rooms. Also, musical instruments can be reverberating. A guitar consists of the top and back plates, ribs, neck, etc., where the impulses of the string acting on the top plate are reflected over the instrument's boundaries, leading to a reverberated and filtered version of the string sound. Piano soundboards are reverberating plates with massive damping\cite{baderpiano}. So the calculated echodensity is non-zero already with a dry musical signal and expected to be used by listeners in spatial sound judgments. The reverberation added alters the echodensity, according to the reverberating space, which adds spaciousness to the perception.

This parameter is not commonly used, but will turn out to be of great importance in room acoustic perception, discussed below.

\section{Results}
\label{results}

\subsection{Acoustic and Psychoacoustic Parameters}
Table \ref{tab:psysri} shows the magnitude of the acoustic and psychoacoustic measures of the impulse responses. As expected, they do not vary that much, as we only compare $9$ impulse responses from $3$ concert venues. 

\begin{table*}
    \centering
    \begin{tabular}{c|c|c|c|c|c|c|c|c|c|c|c|c}
         Stimulus & N & S  & R$_S$ & R$_B$ & F & K & C & RT$_{30}$  & RT$_{20}$ & EDT & E & D\\\hline
         F7& 54.35 & 2.58500& 0.09155& 0.00068 & 0.06925 & 0.2955&1978&1.26244&1.29936&1.25011&0.04402&3.89782\\\hline      
         F10&53.30&2.49000 &0.09030 &0.00067 &0.05445 &0.3285 & 2040&1.20072&1.21364&1.38692&0.04318&3.74436\\\hline				
         F13&55.75&2.76500&0.08195&0.00072&0.08315&0.3020 &1995&1.16547&1.20247&1.16653&0.04437&4.02064\\\hline         
         M3&57.50&2.85000&0.08385&0.00073&0.05965&0.2825 &2153&1.42086&1.44361&1.40800&0.04421&4.30424\\\hline
         M8&53.45&2.62500&0.08275&0.00061&0.04755&0.2590 &2076&1.44211&1.56659&1.61131&0.04364&4.33177\\\hline
         M12&58.00&2.91000&0.08380&0.00082&0.04755&0.2585&2007&1.46840&1.43233&1.62089&0.04599&4.47511\\\hline
         L6&60.50&2.89500&0.08840&0.00097&0.03765&0.2805 &2173&1.96744&1.94750&2.00581&0.04429&4.35494\\\hline
         L7&60.20&2.72000&0.08670&0.00114&0.04180&0.2490&1713&1.91874&1.91931&2.02114&0.04371&4.18609\\\hline
         L13&53.70&2.74000&0.07910&0.00058&0.05665&0.2835 &2087&1.91925&1.94642&1.81592&0.04439&4.25280\\\hline
    \end{tabular}
    \caption{Acoustic and psychoacoustic measures of the $9$ room impulse responses. Concert spaces are F = Fabrik, M = Markthalle, L = Leiszhalle, numbers are internal IRs taken.}
    \label{tab:psysri}
\end{table*}

\subsection{Perceptual proximity ratings}
%
All $7$ values from the Likert scale have been used by the participants, with a median of $4.09$ mean value of $4.19$, a standard deviation of $0.915$. 

\subsection{Multi Dimensional Scaling}

\begin{figure}[ht]
\includegraphics[width=\linewidth]{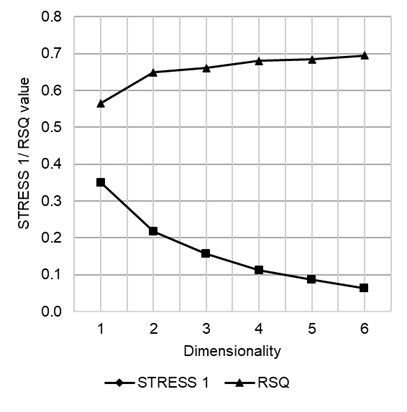}
\caption{Stress 1 and RSQ for MDS embedding dimensions $1$ to $6$.}
\label{fig:Stress_RSQ}
\end{figure}

The results of the MDS analysis of the proximity ratings have been aggregated according to STRESS 1 and RSQ of six MDS evaluations, starting from a one-dimensional up to a six-dimensional embedding,  summarized in Table \ref{tab:MDS_Embeddings} and plotted in Fig. \ref{fig:Stress_RSQ}.

\begin{table}
    \centering
    \begin{tabular}{ccccccc}
        MDS Solution & 6& 5& 4& 3& 2& 1\\
        Stress 1 &$0.06263$	&$0.08623$ & $0.11206$ & $0.15687$ & $0.21715$ & $0.34853$ \\
        RSQ & $0.69443$ & $0.68375$ & $0.68029$ & $0.66067$ & $0.64841$ & $0.56490$\\
    \end{tabular}
    \caption{Stress 1 and RSQ values of the six MDS embeddings.}   \label{tab:MDS_Embeddings}
\end{table}

Fig. \ref{fig:Stress_RSQ} exhibits a gradual decrease in slope of STRESS 1. Only the $4$- to $6$-dimensional solutions exhibit a fair to good STRESS 1, according to Table \ref{tab:Model_Quality}, and each dimension sufficiently reduces STRESS 1 further. So only the models with four to six dimensions are analyzed, 

It is interesting to see that for a fair interpretation at least four dimensions are necessary while a good fit needs six. This is in contrast to timbre perception with a good fit normally already present with three dimensions, as discussed above.

The next step is to interpret the perceptual dimensions found. All nine sounds align in a certain order and position along all dimensions found. The task is, therefore, to correlate these orders and positions of the sounds on each dimension with all acoustic and psychoacoustic parameters calculated from the sounds. The parameters with reasonable correlations then allow for interpretation of the respective perceptual dimension as caused by the best-correlating acoustic or psychoacoustic parameter.

\subsection{Interpretation of the MDS Dimensions}
\begin{figure}[ht]
\includegraphics[width=\linewidth]{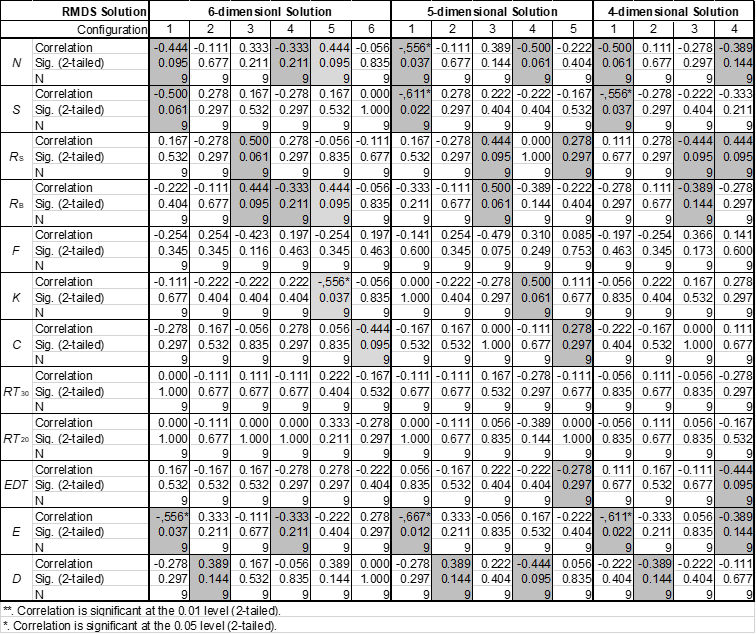}
\caption{\label{pic:correlationmatrix}{Correlation Matrix of the MDS dimensions with the psychoacoustic parameters for 4-, 5-, and 6-dimensional MDS embeddings.}}
\label{pic:tab}
\end{figure}

To interpret the MDS dimensions, we correlate them with the  acoustic and psychoacoustic parameters. As recommended in \cite[p. 344 and p. 353]{field}, we use Kendall's rank correlation, since the data exhibits no normal distribution. Correlations between parameters and dimensions are listed in Table \ref{pic:tab}.

The correlations considered important are visually highlighted. For each embedding, up to three parameters with the highest correlation are considered as potential explanatory candidates. These values are colored gray, where the values in the fifth and sixth dimension of the $6$-dimensional solution are marked with a lighter shade of gray, since these two dimensions were not considered further, as will be explained.

\subsubsection{Dimension 1}
The three acoustic parameters \emph{loudness} $N$, \emph{sharpness} $S$, and \emph{echo density} $E$ tend to correlate with the first dimension of each MDS solution. The strongest and most significant parameter is the echo density. It is negatively related to the first dimension, and the correlation coefficients lie in the range of a medium to strong relation ($t_{b} =-0.611$, $p=0.022$ (4-dimensional), $t_{b}=-0.667$, $p=0.012$ (5-dimensional), $t_{b}=-0.556$, $p=0.037$ (6-dimensional)).

\subsubsection{Dimension 2}
In all MDS solutions, the second dimension correlates strongest with the fractal correlation dimension $D$. It is therefore the most explanatory measure, despite its comparably weak correlation ($t_b=-0.389$, $p=0.144$ (4-dimensional), $t_b=0.389$, $p=0.144$ (5-dimensional), $t_b=0.389$, $p=0.144$ (6-dimensional)).

\subsubsection{Dimension 3}
The third dimension exhibits weak to medium  correlations with roughness according to Sottek $R_\mathrm{S}$ ($t_{b}=-0.444$, $p=0.095$ (4 dimensional), $t_{b}=0.444$, $p=0.095$ (5 dimensional), $t_b=0.500$, $p=0.061$ (6 dimensional)) and according to Bader ($R_\mathrm{B}$) ($t_{b}=-0.389$, $p=0.144$ (4 dimensional), $t_{b}=0.500$, $p=0.061$ (5 dimensional), $t_{b}=0.444$, $p=0.095$ (6 dimensional)). 

\subsubsection{Dimension 4}
\begin{figure}
    \centering \includegraphics[width=\linewidth]{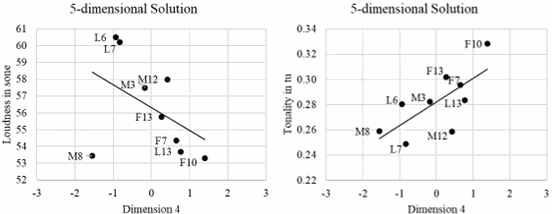}
    \caption{Loudness (left) and tonality (right) over dimension $4$ in the 5 dimensional MDS. Tonality is in much better agreement.}
    \label{pic:5dim4}
\end{figure}

With regard to the fourth dimension, the situation is less clear. In the 4 dimensional MDS, the highest correlation is with Early Decay Time (EDT) ($t_{S}=0.444$, $p=0.095$). In the 5 dimensional solution, both loudness ($N$) and tonality ($K$) correlate with this dimension ($t_{S}=0.500$, $p=0.061$). A visual inspection of parameter over dimension, as plotted in Fig. \ref{pic:5dim4}, is clearly in favor of tonality.

\begin{figure}
    \centering \includegraphics[width=\linewidth]{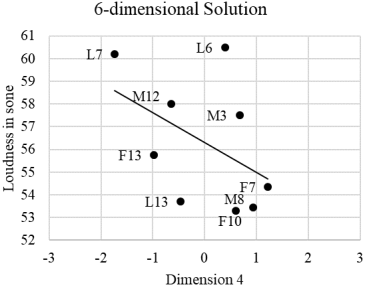}
    \caption{Loudness over dimension $4$ in the 6 dimensional MDS.}
    \label{pic:6dim4}
\end{figure}

In the 6 dimensional MDS, this dimension correlates equally well with loudness ($N$), roughness ($R_S$), and echo density ($E$) ($t_{b}=0.333$, $p=0.211$). As echo density is already considered explanatory for dimension $1$, and roughness explains dimension $3$, loudness is the natural choice. However, as one can see in Fig. \ref{pic:6dim4}, the relationship is not that strong.

\subsubsection{Dimension 5}
In the 5-dimensional solution Roughness ($R_\mathrm{S}$), the spectral centroid ($C$), and Early Decay Time (EDT) show the strongest, but fairly weak correlation with dimension 4 ($t_{S}=0.278$, $p=0.298$). Roughness is already associated with dimension $3$, leaving spectral centroid and early decay time as potential candidates.

\begin{figure}
\centering \includegraphics[width=\linewidth]{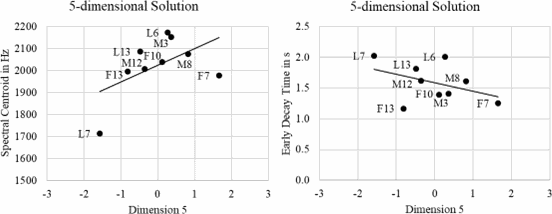}
\caption{Loudness (left) and tonality (right) over dimension $4$ in the 5 dimensional MDS. Tonality is in much better agreement.}
\label{pic:5dim5}
\end{figure}

Figure \ref{pic:5dim5} shows $C$ and EDT over dimension 5. Both seem to be an equally good fit. However, when we consider item L7, an outlier, the correlation between $C$ and dimension 5 vanished completely ($t_{b} =0.071$, $p=0.805$), leaving EDT the only plausible parameter to explain dimension 5.

In the 6 dimensional solution, dimension $5$ correlates with tonality $K$ ($t_\text{b}=0.556$, $p=0.037$). This is plausible, as tonality was also able to explain the fourth dimension in the 5-dimensional solution.

\subsubsection{Dimension 6}
Dimension $6$ correlates with spectral centroid $C$ ($t_\mathrm{b}=0.444$, $p=0.095$). Again, this is plausible, as $C$ is not associated with the other domains.

\subsection{Selection of the Final MDS Solution}
Due to the lack of interpretability of the fourth dimension, the $6$-dimensional MDS solution had to be excluded from further analysis. Accordingly, only the $5$-dimensional and $4$-dimensional solution were available as possible final solutions.

In summary, the echo density $E$ can basically be assigned to the first dimension, as it is significantly correlated to this dimension in the $5$- and $4$-dimensional solution. Although the dimension also correlates relatively strongly and in part also significantly with the two parameters sharpness and loudness, it was decided to maintain the echo density as explanation for the first dimension based on the higher 
$b$-values and significance values that were calculated in terms of this parameter.

The results for the second dimension can be summarized such that only the fractal correlation dimension $D$ could be considered an appropriate explanation for the dimension in the $5$- and $4$- dimensional MDS solution. The reason for this is that in both solutions, the strongest correlations could be found regarding this parameter. However, the correlations were not significant.

As with the first and second dimension, a general statement based on the $5$- and $4$-dimensional solution could be made for the third dimension. In both solutions, either the roughness according to Sottek RS or the roughness according to Bader RB could be assigned to the dimension, where Kendall’s tau-b was used as the decision-making basis. In the $5$-dimensional solution, the roughness according to Sottek correlates more strongly with the third dimension, while in the $4$-dimensional solution the roughness according to Bader correlates more strongly with it. However, highly significant relationships could not be established in either case. But still, it was not possible to say which of the two roughness models should be assigned to the third dimension because it was not yet clear which MDS solution should be chosen as the final solution. For this, first of all, the fourth and fifth dimensions had to be considered in order to find out how many dimensions should ultimately be considered regarding the perception of reverberance when listening to music. Consequently, it had to be discussed which parameter should be used as an explanation for the fourth or, if applicable, for the fifth dimension.

Overall, it had been found that in both the $5$- and $4$-dimensional solution the loudness N could be used as a possible explanation for the fourth dimension, whereby with regard to the $4$-dimensional solution, the EDT explained the dimension even better. Therefore, with respect to the $4$-dimensional solution, the EDT and, with respect to the $5$-dimensional solution, the loudness was assigned to the fourth dimension. 

At this point, however, it is worth taking a look at the fifth dimension as well because the EDT can also be used as an explanation here. Thus, while in the $5$-dimensional solution the loudness can be assigned to the fourth dimension and the EDT can be assigned to the fifth dimension, in the $4$-dimensional solution only the EDT comes into question as a further perceptual dimension.

Consequently, by choosing the $5$-dimensional solution as the final solution, more dimensions of the perceptual space could be explained, which would make sense for the following reason. When the first dimension is considered again, it can be noticed that there is also a moderate and, in some cases, even significant relationship between the loudness $N$ and the dimension in all relevant MDS solutions. Accordingly, it is reasonable to suppose that loudness is in some way related to the perception of reverberance, too. However, to be able to represent this relationship, the parameter must be assigned to one dimension individually, whereby the choice of the $5$-dimensional MDS solution would allow this assignment as previously explained. 

Consequently, the $5$-dimensional solution must be selected as final MDS solution. Moreover, the choice of the $5$-dimensional solution leads to the third dimension being assigned the roughness according to Bader $R_\text{B}$.

Thus, as a final result, the perception of reverberance when listening to music could be related to the parameters of Table \ref{tab:5-dim_Space}.

\begin{table}
    \centering
    \begin{tabular}{c|c}
        Dimension & Parameter \\ \hline
        $1$ & Echodensity ($E$)\\ \hline
        $2$ & Fractal Correlation Dimension ($D$)\\\hline
        $3$ & Roughness (Helmholtz/Bader) (RB\\\hline
        $4$ & Loudness ($N$)\\\hline
        $5$ & Early Decay Time (EDT)\\
    \end{tabular}
    \caption{Final $5$-dimensional perception space of spatial perception.}
    \label{tab:5-dim_Space}
\end{table}


Interestingly, the reverberation times $\text{TR}_{20}$ and $\text{TR}_{30}$ exhibit no correlation with the perceptual dimensions of subjective room acoustics, in contrast to early decay time. This underlines the claim that the early decay time is better suited to predict subjective room acoustics (see, e.g., \cite{ziemerbook}).

\section{\label{sec:5}Conclusion}
In this study, we presented a Multi Dimensional Scaling approach to study subjective room acoustics. Furthermore, we introduced the echo density ($E$) and the fractal correlation dimension ($F$) as potential explanatory parameters. 

It appears that the room acoustical impression is four to six dimensional. This finding validates the assumption that subjective room acoustics are multidimensional \cite{Beranek1962,saqi}. These perceptual dimensions are associated with echo density, fractal correlation dimension, roughness and, less distinctly, with Early Decay Time, tonality and spectral centroid. These parameters deviate a lot from the usual suspects. It is possible that conventional approaches with semantic attributes and bipolar adjective scales produce a bias that vanished due to the MDS approach. MDS methodology avoids all verbalization, eliminating such a bias.

Echo density and fractal correlation dimension explain the first and second MDS dimension, i.e., the most relevant ones. This observation underlines that they are meaningful acoustic parameters that should be consulted more often in room acoustical studies. Further examination is necessary to reveal their relationship with auditory perception. Lastly, the results underline the idea that early decay time is in better agreement with the perception of reverberation than the reverberation times ($\text{RT}_{20}$ and $\text{RT}_{30}$).

Repeating the study with other participants, impulse responses, and/or convolved music pieces will help examine the reliability of the findings.


\section*{Author Declarations}
We have no conflicts of interest to disclose. Experiment data and audio material can be sent on request.

\appendix


\bibliography{sampbib}

\end{document}